\documentclass[preprint,notoc]{JHEP3}
\usepackage{epsfig}
\def\eslt{\not\!\!{E_T}}
\def\emiss{\not\!\!{E}}

\def\to{\rightarrow}

\def\bi{\begin{itemize}}
\def\ei{\end{itemize}}
\def\be{\begin{equation}}
\def\ee{\end{equation}}
\def\bea{\begin{eqnarray}}
\def\eea{\end{eqnarray}}
\def\te{\tilde e}

\def\ttau{\tilde \tau}
\def\tmu{\tilde \mu}
\def\tg{\tilde g}

\def\tw{\widetilde W}
\def\tz{\widetilde Z}
\def\alt{\stackrel{<}{\sim}}

\def\shat{\hat{s}}
\title{
Two Photon Background and the Reach\\ 
of a Linear Collider for Supersymmetry\\
in WMAP Favored Coannihilation Regions
}

\author{Howard Baer and Tadas Krupovnickas
\\ Department of Physics, Florida State University\\ 
Tallahassee, FL 32306, USA\\
E-mail: \email{baer@hep.fsu.edu}, \email{tadas@hep.fsu.edu}}

\author{Xerxes Tata
\\ Department of Physics, University of Hawaii\\ 
Honolulu, HI 96822, USA\\
E-mail: \email{tata@phys.hawaii.edu}}

\preprint{\vbox{\hbox{FSU-HEP-040505}\vspace{0.2cm}\hbox{UH-511-1050-04}}}

\abstract{
A neutralino relic density in accord with WMAP measurements can be found
in the minimal supergravity (mSUGRA) model in several regions of
parameter space:
the stau co-annihilation corridor at low $m_0$ and the 
hyperbolic branch/focus point (HB/FP) region at large $m_0$ at
the edges of parameter space, and the bulk and Higgs boson funnel
regions within. 
In the regions at the edge, the mass gap between
the next-to-lightest SUSY particle (NLSP) and the lightest SUSY particle (LSP)
becomes small, and backgrounds from $\gamma\gamma\to f\bar{f}$ 
($f$ is a SM fermion) become important for NLSP detection at an $e^+e^-$
linear collider.
We evaluate these backgrounds from bremsstrahlung and beamstrahlung photons, 
and demonstrate that these do not preclude the observability of the signal 
for the cases where either the stau or the chargino
is the NLSP.
We also delineate the additional portion of
the stau coannihilation region, beyond what can be accessed via
a search for selectrons and smuons, that can be probed 
by a search for
di-tau-jet events plus missing energy.
The reach of a LC for SUSY in the HB/FP region is shown 
for an updated value of $m_t\simeq 180$ GeV as recently 
measured by the $D\O$ \ experiment.
}

\keywords{Supersymmetry Phenomenology, e+e- Experiments, %
Dark Matter, Supersymmetric Standard Model}

\begin{document}

\section{Introduction}
\label{sec:intro}
%

Weak scale supersymmetry is a highly motivated extension of the
Standard Model (SM)\cite{review}.
In models with gravity-mediated supersymmetry breaking and $R$-parity
conservation, the lightest SUSY particle is usually the lightest 
neutralino $\tz_1$, which is absolutely stable and serves as a good
candidate particle for cold dark matter (CDM) in the universe\cite{haim}.
Indeed, the recent precision mapping of anisotropies in 
the cosmic microwave background radiation by the 
WMAP collaboration\cite{wmap,dimitri}
has allowed a determination of
\be
\Omega_{CDM}h^2 =0.1126^{+0.0161}_{-0.0181}
\ee
at the $2\sigma$ level, where $\Omega =\rho_{CDM}/\rho_c$ is the ratio
of the cold dark matter density to the critical mass density of the
universe, and $h$ is the scaled Hubble constant.  This DM is unlikely to
all be massive black holes (at least in our galactic halo)
\cite{macho}, or ordinary baryons (in the form of brown dwarfs) since a
baryonic density of this magnitude would lead
to conflicts with both Big Bang nucleosynthesis and the 
density of baryons as determined from the acoustic peaks
in the CMB spectrum.

The minimal supergravity model (mSUGRA) forms a 
convenient template for exploring the experimental consequences of
gravity-mediated SUSY breaking models\cite{msugra}.
The mSUGRA model is characterized by four SUSY
parameters together with a sign choice,
\be
m_0,\ m_{1/2},\ A_0,\ \tan\beta\ \ {\rm and}\ sign(\mu ).
\ee
Here $m_0$ is the common mass parameter
for all scalar particles at $Q=M_{GUT}$, $m_{1/2}$ is the common 
gaugino mass at $M_{GUT}$, $A_0$ is the common trilinear soft term at
$M_{GUT}$, $\tan\beta$ is the ratio of Higgs field vacuum expectation 
values at the scale $M_Z$, and
finally the magnitude -- but not the sign -- of the superpotential
$\mu$ term is determined by the requirement of radiative electroweak 
symmetry breaking (REWSB). In addition, the top quark mass
$m_t$ must be specified. While the PDG quotes a value $m_t=174.3\pm 5.1$
GeV\cite{pdg}, we note that the $D\O$ \ experiment has recently determined
a value $m_t=180.1\pm 5.1$ GeV by a re-analysis of 
Run 1 data\cite{d0mt}.

Over the years, several groups have evaluated the relic density of
neutralinos in the context of the 
mSUGRA model\cite{Afunnel,stau}. This has been recently 
re-examined 
in light of the WMAP data\cite{wmap_pap}.
A value of $\Omega_{\tz_1}h^2$ in accord with the WMAP
determination can be found
in four broad regions of model parameter
space:\footnote{We note that the CDM may consist of
  several components, so that WMAP observation should be interpreted as
  an upper bound $\Omega_{\tz_1}h^2 < 0.129$ on the neutralino relic
  density. If we make the additional assumption that relic neutralinos
  consititute all the CDM, more stringent constraints are possible.}  
\begin{itemize}
\item The bulk annihilation region occurs at low $m_0$ and low $m_{1/2}$
for all $\tan\beta$ values. In this region, neutralino annihilation
in the early universe occurs dominantly via $t$-channel exchange of light 
sleptons. 
\item The stau co-annihilation region occurs at low $m_0$ 
where $m_{\ttau_1}\simeq m_{\tz_1}$, and where
$\ttau_1 -\tz_1$ and $\ttau_1-\bar{\ttau}_1$ 
annihilation in the early universe serve to 
reduce the neutralino relic density to sufficiently low values\cite{stau},
\item the hyperbolic branch/focus point region\cite{ccn,fmm,bcpt} (HB/FP) at
large $m_0$ near the edge of parameter space where $\mu$ becomes
small, and the $\tz_1$ has a significant higgsino component which
facilitates a large annihilation rate\cite{bb2,fmw,bb}. The location of
this region is very sensitive to the value of $m_t$ \cite{bkt}.
\item the $A$-annihilation funnel at large $\tan\beta$ where
$m_H$ and $m_A\simeq 2m_{\tz_1}$ and $\tz_1\tz_1\to A,\ H\to f\bar{f}$ 
($f$'s are
SM fermions) through the very broad $A$ and $H$ 
resonances\cite{Afunnel}.
\end{itemize}
The bulk annihilation region is now disfavored and possibly excluded
because it gives rise to values of $m_h$ in violation of LEP2 limits,
and can give large deviations from $BF(b\to s\gamma )$ and
$(g-2)_\mu$\cite{constraints}.  A portion of the bulk annihilation
region may survive, but only where it overlaps with the stau
co-annihilation region or the $A$ annihilation funnel. Other regions of
parameter space may also give a reasonable relic density, including the
stop co-annihilation region (for very particular $A_0$
values)\cite{stopco} and the $h$ annihilation corridor, where
$2m_{\tz_1}\simeq m_h$.

Once the dark matter allowed regions of mSUGRA parameter space are
identified, it is useful to see what the implications of these regions
are for collider experiments. The reach of the Fermilab
Tevatron\cite{bkt,earlier} and CERN LHC\cite{lhc} have recently been
worked out for the DM allowed regions of parameter space. In addition,
the reach of a linear $e^+e^-$ collider has also been investigated for
the DM allowed regions of the mSUGRA model\cite{bbkt}.

In Ref. \cite{bbkt}, it was found that a $\sqrt{s}=500-1000$ GeV LC
could probe the entire stau co-annihilation region for $\tan\beta \alt 10$
by searching for selectron and smuon pair events. For higher $\tan\beta$
values, the stau co-annihilation region increases in $m_{1/2}$, 
while the reach for dilepton pairs decreases. The decrease in dilepton
reach occurs mainly because the large-reach region is subsumed by the
unallowed region where the $\ttau_1$ becomes the LSP (a $\ttau_1$ LSP
is not allowed in $R$-conserving models because it would lead to charged stable
relics from the Big Bang, for which there are stringent search limits).
Some additional reach may be gained in the stau co-annihilation region by
searching for stau pair production events, although this possibility 
was not investigated in Ref. \cite{bbkt}.

The reach of a LC for SUSY in the HB/FP region was also examined in
Ref. \cite{bbkt}. It was found that in this region a chargino pair
production signal could be seen above SM background essentially up to
the kinematic limit for production of two charginos at the LC. Standard
cuts for identifying $\ell+2-jet$ events allowed chargino detection
in the low $m_{1/2}$
portion of the HB/FP region. In the high $m_{1/2}$ portion of
the HB/FP region, the $\tw_1 -\tz_1$ mas gap became so small that
specialized cuts were needed to pick out the signal from the
background. In addition to usual SM backgrounds from
the production of heavy particles ($W$, $Z$, $t$), contributions
to backgrounds 
from $e^+e^-\to f\bar{f}$
and $\gamma\gamma\to f\bar{f}$ ($f=b,c$) were also included.
The two photon cross sections
were evaluated using Pythia\cite{pythia} for the
bremsstrahlung photon distribution, while the beamstrahlung contribution
was not included. It was found that in the HB/FP region, an $e^+e^-$ LC
with $\sqrt{s}=500-1000$ GeV could have a reach for SUSY considerably
beyond that of the LHC!

It is significant that a neutralino relic density in accord with WMAP analyses
occurs along  the boundary of parameter space for both the
stau co-annihilation case and the HB/FP region. In both these cases, 
the next-to-lightest SUSY particle (NLSP) becomes nearly degenerate
in mass with the LSP. Thus, although NLSPs may be produced at large 
rates at colliders, 
$NLSP\to LSP$ decays may result in only a small visible energy release, 
potentially making their detection difficult.
This is because $\gamma\gamma \to f\bar{f}$
production, where the photons may originate from bremsstrahlung 
off the initial state electrons, or via beamstrahlung constitute
an important background to the signal. 
In the stau co-annihilation region, we do not expect the two photon background
to be problematic for the search for selectron or smuon pairs. The reason is
that the $e^+e^-$ or $\mu^+\mu^-$ pair originating from $\gamma\gamma$
annihilation is expected to come out back-to-back in
the transverse plane for the background, but not so for the SUSY signal.
However, for stau pair production, the $\gamma\gamma \to \tau\bar{\tau}$
background can be important, since the tau lepton is unstable, and the
visible tau decay products will in general not be back-to-back in the 
transverse plane, especially when the daughter taus are relatively soft.
 
One of the purposes of this paper
is a careful examination of the two photon background, including
both bremsstrahlung and beamstrahlung contributions\cite{drees_god}. 
We describe our treatment of the $\gamma\gamma\to f\bar{f}$
background , and its inclusion 
into Isajet v7.70\cite{isajet} in Sec.~\ref{sec:2photon}.
We note here that it is likely that far forward detector elements
reaching as close as 25 mrad of the beam line will likely be
implemented in a realistic detector. The main purpose of the 
forward detectors will be to veto $\gamma\gamma$ or $\gamma e$ events
where the initial $e^\pm$ suffers a slight deflection into the
instrumented regions. In this work, we operate under the assumption that
the initial state $e^\pm$ are collinear with the beam. In this sense, 
our results may be regarded as conservative. 

In Sec. \ref{sec:stau}, we examine extending the LC reach for SUSY in
the stau co-annihilation region by searching for signals with two tau
jets plus missing energy\cite{dutta}.  The additional reach gained
beyond that from the di-electron or di-muon channel is in fact
substantial, especially for large $\tan\beta$. In Sec. \ref{sec:hb_fp},
we re-examine the reach for SUSY by a LC in the HB/FP region, including
the $\gamma\gamma \to b\bar{b}$ and $c\bar{c}$ background to $\ell
+2$-jet events originating from both bremsstrahlung and beamstrahlung
processes.  The added background from beamstrahlung is substantial, but
can be eliminated by invoking an additional angular cut in addition to
cuts already suggested in Ref. \cite{bbkt}.  We present updated reach
results for one value of $\tan\beta$ at a higher value of $m_t=180$ GeV
as suggested by a recent D\O \ analysis\cite{d0mt}.
In Sec. \ref{sec:conclude} we present a summary and conclusions of
our results.

\section{Two photon backgrounds}
\label{sec:2photon}

Our goal in this section is to evaluate the $e^+e^-\to f\bar{f}$
background including contributions from bremsstrahlung and beamstrahlung,
and including two photon annihilation processes.
In this case, the $e^+e^-\to f\bar{f}$ cross section is given by
\bea
\sigma (e^+e^-\to f\bar{f})=\int dx_1 dx_2\left[ f_{e/e}(x_1,Q^2)
f_{e/e}(x_2,Q^2)\hat{\sigma}(e^+e^-\to f\bar{f})\right.\nonumber\\
+\left.
\left( f_{\gamma /e}^{brem}(x_1,Q^2 )+f_{\gamma /e}^{beam}(x_1)\right)
\left( f_{\gamma /e}^{brem}(x_2,Q^2 )+f_{\gamma /e}^{beam}(x_2)\right)
\hat{\sigma} (\gamma\gamma\to f\bar{f})\right] .
\eea
Here, $f_{e/e}(x,Q^2)$ is the parton distribution function 
(PDF) for finding an electron inside the electron, and is given by
the convolution
\be
f_{e/e}(x,Q^2)=\int_x^1 dz f_e^{brem}(\frac{x}{z},Q^2)f_e^{beam}(z)/z ,
\ee
where $f_e^{brem}(x,Q^2)$ is the bremsstrahlung parton distribution 
function of the electron, and $f_e^{beam}(x)$ is the beamstrahlung
parton distribution function of the electron.
We use the Fadin-Kuraev distribution function for
bremstrahlung electrons\cite{fk}, given by
\be 
f_{e/e}^{brem}(x,Q^2)=\frac{1}{2}\beta \left(1-x\right)^{\frac{\beta}{2}-1}
\left(1+{3\over 8}\beta\right)-{1\over 4}\beta \left(1+x\right) ,
\ee
where $\beta =\frac{2\alpha}{\pi}(\log\frac{Q^2}{m_e^2} -1)$.  For the
beamstrahlung distribution function of electrons in the electron, we use
the parametrization of Chen\cite{chen}, which is implemented in terms of
the beamstrahlung parameter $\Upsilon$ (which depends in various
characteristics of the beam profile), beam length $\sigma_z$ (given in
mm), and the beam energy $E_e$.  For $f_{\gamma /e}^{brem}(x,Q^2)$,
the bremsstrahlung distribution function of photons in the electron, we
use the Weizacker-Williams distribution. Finally, 
$f_{\gamma /e}^{beam}(x)$,
the beamstrahlung distribution function of photons inside the
electron, is related to $f_{e /e}^{beam}(x)$, 
and is again determined by the
values of $\Upsilon$, $\sigma_z$ and $E_e$~\cite{chen}.

As an illustration, we plot in 
Fig. \ref{fig:brbeam} the electron bremsstrahlung distribution function
(dashed curve), the electron beamstrahlung distribution (dot-dashed curve)
and the convolution (solid curve). For the  beamstrahlung function, we take
$E_e=250$ GeV, $\Upsilon=0.1072$, and $\sigma_z =0.12$ mm, 
typical for a $\sqrt{s}=500$ GeV $e^+e^-$ LC using
X-band klystron accelerating technology.
We see that the distribution of electrons within the 
electron is strongly peaked at $x\simeq 1$, and that there is a
long tail extending to low fractional momentum $x$.

%
\FIGURE{\epsfig{file=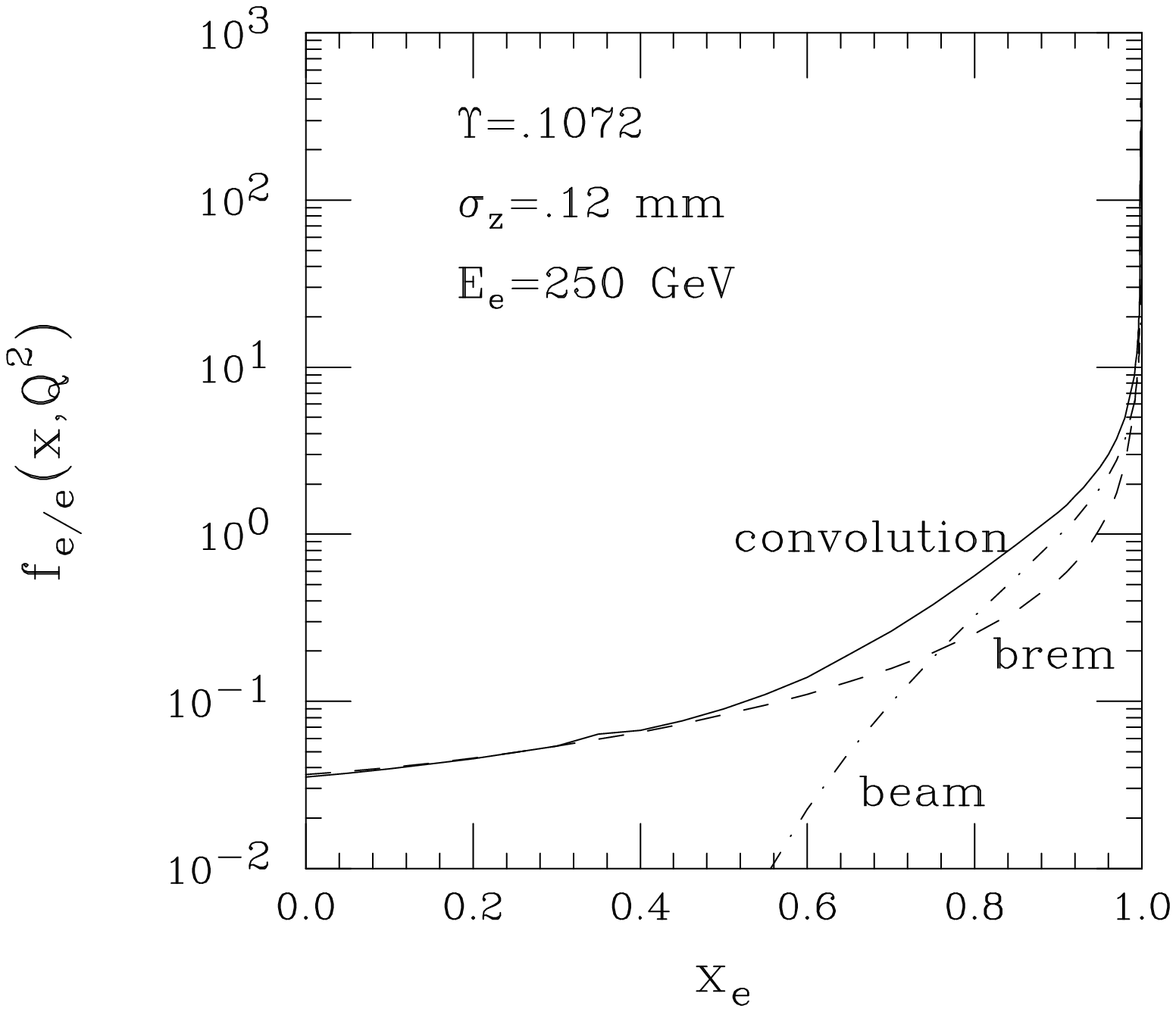,width=12cm} 
\caption{Electron parton distribution functions in the 
electron from bremsstrahlung and from beamstrahlung, along with
their convolution, 
for $E_{beam}=250$ GeV, $\Upsilon=0.1072$ and $\sigma_z =0.12$ mm.
}
\label{fig:brbeam}
}

In Fig. \ref{fig:beam}, we show the corresponding bremsstrahlung (solid)
and beamstrahlung (dashed) distribution functions for photons inside the
electron, for the same beam parameters as in Fig. \ref{fig:brbeam}.  In
this case, both distribution functions are sharply peaked at $x\sim 0$,
which indicates the presence of an intense cloud of soft photons
accompanying the beam of electrons in a linear collider.  The
bremsstrahlung photon distribution function remains significant out to
large $x\sim 1$, while the beamstrahlung photon distribution function
(which is anti-correlated with the corresponding electron distribution)
completely dies off for large $x$.
\FIGURE{\epsfig{file=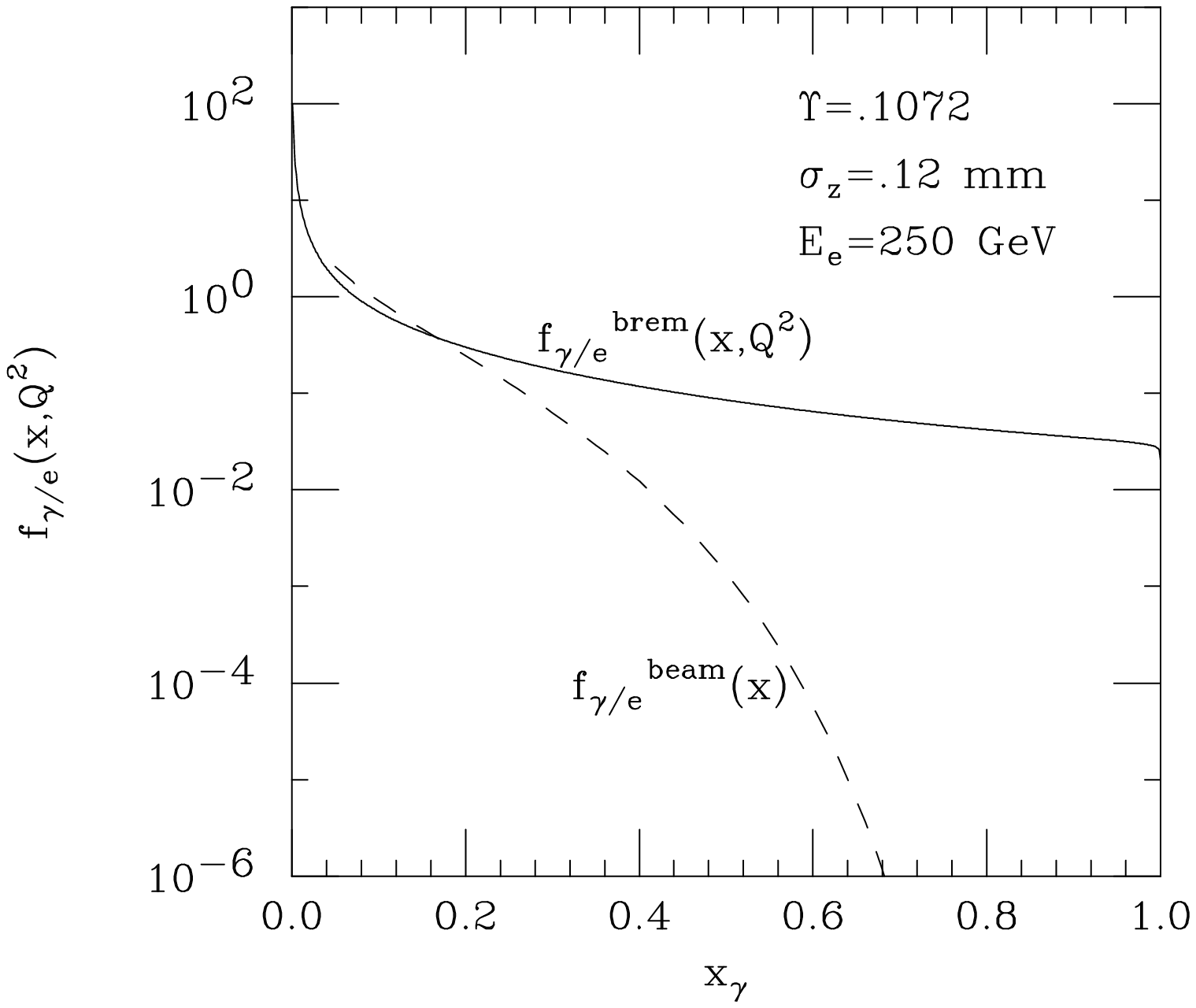,width=12cm} 
\caption{Photon parton distribution functions in the 
electron from bremsstrahlung and from beamstrahlung,
for $E_{beam}=250$ GeV, $\Upsilon=0.1072$ and $\sigma_z =0.12$ mm.
}
\label{fig:beam}
}

The tree level subprocess cross section for $e^+e^-\to f\bar{f}$
via $s$-channel $\gamma$ and $Z$ exchange can be found in many texts
(see for instance Ref. \cite{bp}). The $\gamma\gamma\to f\bar{f}$
subprocess cross section can be easily constructed by crossing,
along with appropriate replacements of electric charge and color
factors from the $e^+e^-\to\gamma\gamma$ cross section given, for instance,
in Ref. \cite{ps}.  
In Fig. \ref{fig:eett}, we show the distribution of 
tau pair invariant mass
for $e^+e^-\to \tau^+\tau^-$ events at a $\sqrt{s}=500$ GeV LC, using the
same beamstrahlung parameters as in Fig.~\ref{fig:brbeam}.
We have required that
$p_T(\tau )>5$ GeV and $|\eta_\tau|<2.5$. The solid histogram represents the
contribution from the $\hat{\sigma}(e^+e^-\to\tau^+\tau^- )$
cross section. The peak at $m(\tau^+\tau^- )=500$ GeV 
originates in the peak of the electron PDF $f_{e/e}(x,Q^2)$ 
at $x =1$. In addition, 
the peak from the radiative return to the $Z$ resonance is clearly 
visible at 
at $m(\tau^+\tau^- )=M_Z$. We note that the enhancement that
would be expected to occur at $x\sim 0$
due to the photon exchange in the subprocess amplitude is eliminated 
by the $p_T$ and $|\eta|$ cuts.
We also plot as the dashed histogram the contribution to
tau pair production from the $\gamma\gamma\to \tau^+\tau^-$
subprocess cross section. It can be seen that this contribution completely 
dominates the tau pair production cross section for invariant masses below
about $\sqrt{\shat}\sim 250$ GeV. In particular, it completely overwhelms
the $Z$ resonance peak, and has the potential to yield a 
formidable background for signal events with only relatively 
soft fermions as visible particles in the final state.

\FIGURE{\epsfig{file=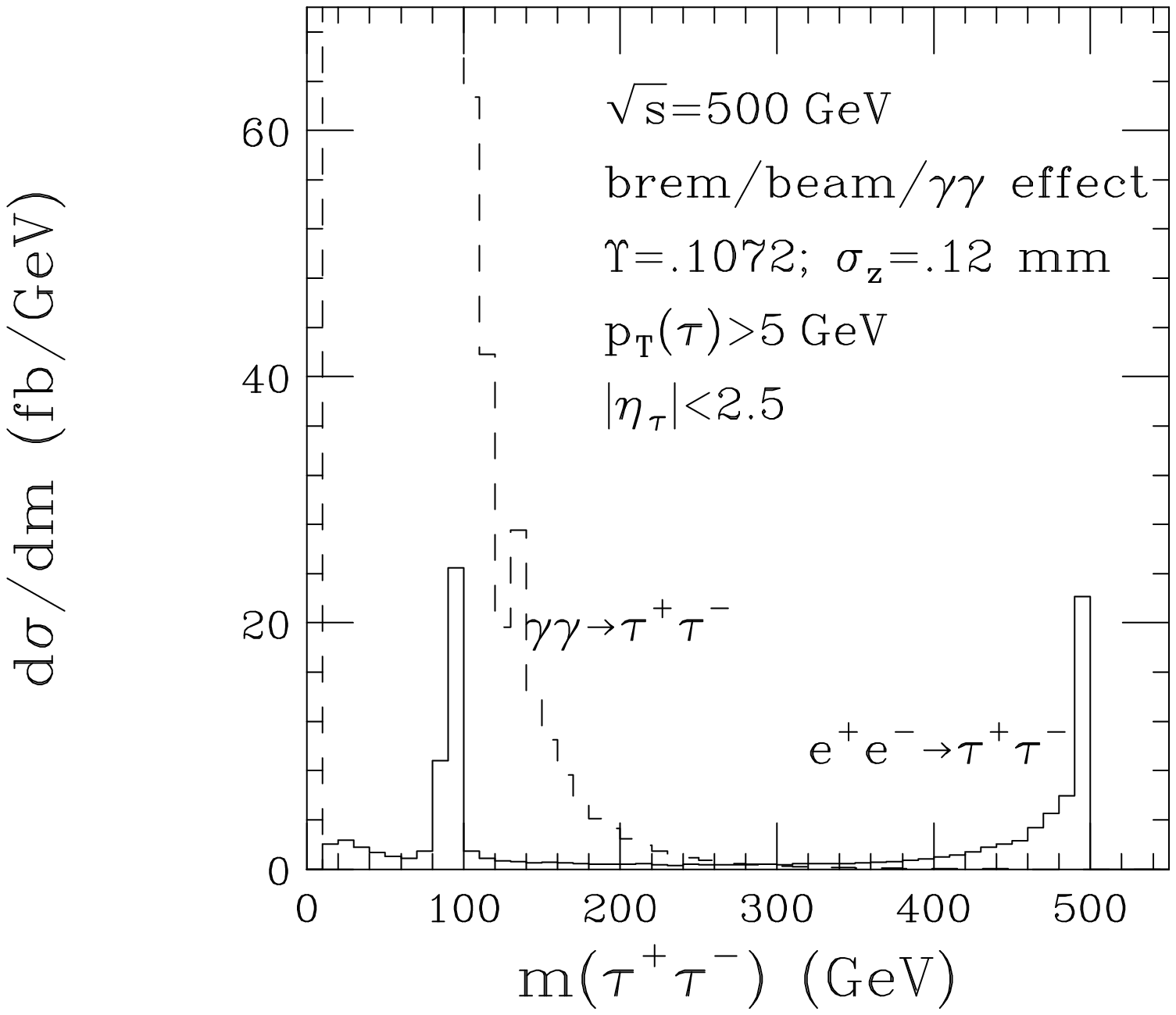,width=14cm} 
\caption{Distribution in di-tau invariant mass from
$e^+e^-\to\tau^+\tau^-$ at a $\sqrt{s}=500$ GeV linear $e^+e^-$ collider
including bremsstrahlung and beamstrahlung effects, and
the two-photon annihilation contribution,
for $E_{beam}=250$ GeV, $\Upsilon=0.1072$ and $\sigma_z =0.12$ mm. Cuts
described in the text have been implemented.
}
\label{fig:eett}
}

The various $\gamma\gamma \to f\bar{f}$ subprocess cross sections
for SM fermions have been incorporated into the event generator 
Isajet v7.70\cite{isajet},
along with the photon bremsstrahlung and beamstrahlung distribution
functions. Electron bremsstrahlung 
and beamstrahlung distributions had already been incorporated previously.

\section{Reach for SUSY in the stau co-annihilation region}
\label{sec:stau}

Our goal in this section is to evaluate the reach of 
an $e^+e^-$ LC for SUSY in the stau co-annihilation region
by searching for stau pair production events.
Most earlier studies
have identified the reach of a LC for SUSY in the
stau co-annihilation region by searching for $e^+e^-$ or
$\mu^+\mu^-$ pairs which would originate from selectron or smuon
pair production\cite{jlc1,bmt,bbkt}; see, however, Ref.~\cite{nojiritau}
for a study of stau signals at the LC. 
In the mSUGRA model at low values of $\tan\beta$, the
$\ttau_1$, $\te_1$ and $\tmu_1$ all tend to be nearly degenerate in mass.
As $\tan\beta$ increases, the $\tau$ Yukawa coupling increases, 
which helps to drive the $\ttau_L$ and $\ttau_R$ soft SUSY breaking masses
to values lower than their first and second generation counterparts.
In addition, if $\tan\beta$ is large, mixing between
$\ttau_L$ and $\ttau_R$ states reduces $m_{\ttau_1}$ even further.
Thus, for low $m_0$ and 
large $\tan\beta$, it turns out that $m_{\ttau_1}$ can be significantly 
lighter than the lightest selectrons or smuons. In this case, there should
exist portions of mSUGRA model parameter space where
$\sigma (e^+e^-\to\ttau_1^+\ttau_1^-)$ is large, while
$\sigma (e^+e^-\to \te_1^+\te_1^-)$ and 
$\sigma (e^+e^-\to\tmu_1^+\tmu_1^-)$
are kinematically suppressed or forbidden. 

For all our computations, we use Isajet 7.70\cite{isajet} which allows
for the use of polarized beams, and also allows for convolution of
subprocess cross sections with electron and photon PDFs to incorporate
initial state bremsstrahlung as well as beamstrahlung effects.  We use
the Isajet toy detector CALSIM with calorimetry covering the regions
$-4<\eta <4$ with cell size $\Delta\eta\times\Delta\phi = 0.05\times
0.05$. Electromagnetic energy resolution is given by $\Delta
E_{em}/E_{em}=0.15/\sqrt{E_{em}}\oplus 0.01$, while hadronic resolution
is given by $\Delta E_h/E_h=0.5/\sqrt{E_h}\oplus 0.02$, where $\oplus$
denotes addition in quadrature, and energy is measured in GeV.  Jets are
identified using the Isajet jet finding algorithm GETJET using a fixed
cone size of $\Delta R=\sqrt{\Delta\eta^2+\Delta\phi^2}=0.6$, modified
to cluster on energy rather than transverse energy.  Clusters with
$E>5$~GeV and $|\eta (jet)|<2.5$ are labeled as jets.  Muons and
electrons are classified as isolated if they have $E>5$~GeV, $|\eta_\ell
|<2.5$, and the visible activity within a cone of $R=0.5$ about the
lepton direction is less than $max(E_\ell/10\ {\rm~GeV},1\
{\rm~GeV})$. Finally, ``$\tau$-jets'' are defined as jets fulfilling the
above jet criteria, but in addition having just one or three charged
tracks included within the jet cone.\footnote{This tends to overestimate
  the chance that a QCD jet fakes a tau jet. We will see, however, that
  our cuts are efficient in removing this non-physics background, and
  the reach in channels involving tau is ultimately limited by the SM
  sources of taus.}

In this section, we assume an integrated luminosity of
100 fb$^{-1}$ for both a $\sqrt{s}=500$ and 1000 GeV $e^+e^-$ LC.
We assume right polarized electron beams with $P_L(e^-)=-0.9$,
to minimize SM background from $W$ pair production, with no cost and even 
a modest enhancement of the stau signal.

We generate SUSY signal events in the mSUGRA model
parameter space, plus all $2\to 2$ SM background processes as incorporated
into Isajet. We then require both signal and background events
with
\begin{itemize}
\item no isolated leptons,
\item two tau jets.
\end{itemize} 
The requirement of two jets restricts $E_{vis}>10$~GeV, as each jet has 
$E>5$~GeV.
At this point, the signal in the stau co-annihilation region
is dominated by tau jet pairs originating from stau pair production,
while background is dominated by $\gamma\gamma\to\tau^+\tau^-$
production. The background $\tau^+\tau^-$ pair comes out 
back-to-back in the transverse plane, and this correlation is largely 
maintained by the visible tau jets. The distribution in $\cos\phi (jj)$,
the dijet opening angle in the transverse plane, is shown
in the first frame of Fig. \ref{fig:dist} for the $\tau^+\tau^-$ background
(red histogram) and all other SM backgrounds ({\it e.g.} $WW$, $ZZ$
production, {\it etc.}) (black histogram). The signal distribution
is shown in the upper right frame, for the mSUGRA point
$m_0=200$ GeV, $m_{1/2}=520$ GeV, $A_0=0$, $\tan\beta =30$ and
$\mu >0$. We also take $m_t=180$ GeV. While the signal has some peak near
$\cos\phi (jj)\sim -1$, it maintains a broad distribution for all other
$\cos\phi (jj)$ values. The $\gamma\gamma\to\tau^+\tau^-$
background is, however, much more sharply peaked at $\cos\phi (jj)\sim -1$, 
which leads us to propose the cut:
\begin{itemize}
\item $\cos\phi (jj)>-0.9$ .
\end{itemize}
\FIGURE{\epsfig{file=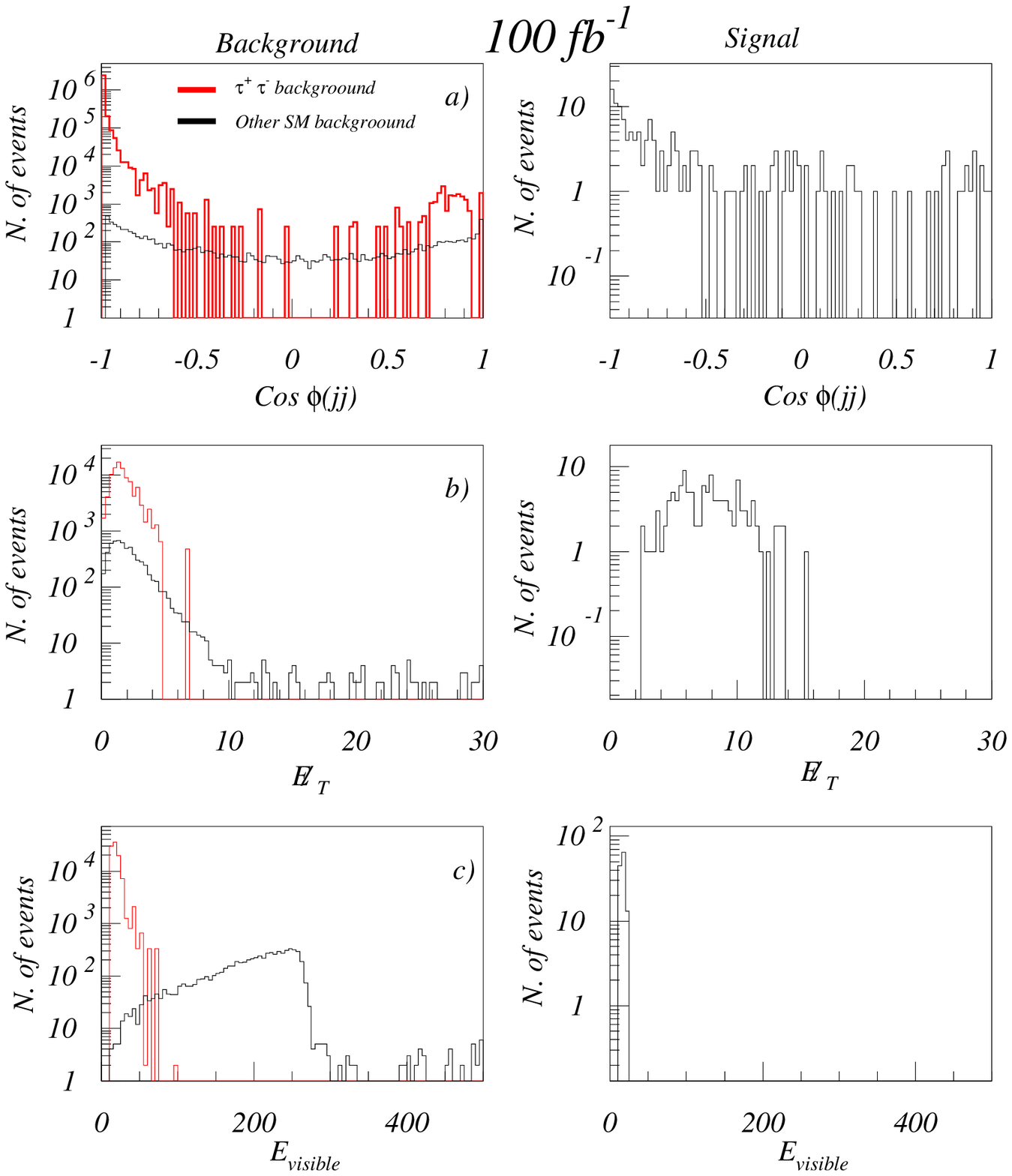,width=14cm} 
\vspace*{-0.8cm}
\caption{Distribution of events in $\cos\phi (jj)$, 
$\eslt$ and $E_{visible}$ for
di-tau jet$+\emiss$ events from 
$\tau\bar{\tau}$ production via both
$e^+e^-$ and $\gamma\gamma$ annihilation, along with
background from all other SM processes (left-hand side).
We also show distributions from stau pair production signal events
at mSUGRA point $m_0,\ m_{1/2},\ A_0,\ \tan\beta,\ sign(\mu )=
200\ {\rm GeV},\  520\ {\rm GeV},\ 0,\ 30,\ +1$, with $m_t=180$ GeV 
for a $\sqrt{s}=500$ GeV LC
(right-hand side).
The relic density for this point is $\Omega_{\tz_1}h^2=0.09$.
}
\label{fig:dist}
}

After the $\cos\phi (jj)$ cut, we show the distributions for
background and signal events in missing transverse energy $\eslt$
(middle two frames) and visible energy $E_{visible}$ (lower
two frames). The optimal cuts in these latter two quantities vary 
depending on where one is in parameter space.
We examine the possibilities:
\begin{itemize}
\item $\eslt >0,\ 5,\ 10, \cdots ,\ 195,\ 200$ GeV, and
\item $E_{visible}<15,\ 20,\ 25,\ \cdots ,\ 495,\ 500$ GeV.
\end{itemize}
For each set of possibilities, we consider only cases 
with at least 10 signal events
for 100~fb$^{-1}$ of integrated luminosity.
We then pick the set of $\eslt,\ E_{visible}$ cuts which maximizes
the $S/(S+B)$ ratio ($S=$ signal and $B=$ background) and at
the same time yields
a $5\sigma$ signal which is our criterion  for observability
against SM backgrounds.\footnote{We recognize that by trying many sets
  of cuts to optimize the signal as a function of parameters, we should
  include a trials factor when assessing the statistical significance of
  the signal. We have not done so. We remark, however, that current
  projections for the expected integrated luminosity of a LC are several
  hundred fb$^{-1}$. Any signal identified by cuts optimized using the
  first 100~fb$^{-1}$ of integrated luminosity, should therefore be visible in
  subsequent runs where the analysis is performed with these same cuts
  so that there is no trials factor for this new analysis.}
For the parameter space point shown in Fig. \ref{fig:dist}, the total
cross section before cuts is
$\sigma(e^+e^-\to\ttau_1^+\ttau_1^-)=17.65$~fb.  The optimal cuts turn
out to be $\eslt >5$ GeV and $10\ {\rm GeV}<E_{visible}<20\ {\rm
GeV}$. Using these cuts, the signal cross section is 0.96 fb, while the
background level is 2.13 fb. We have checked that $S/B \sim 0.3-1$ are
typical.

In Fig. \ref{fig:30p}, we show the $m_0\ vs.\ m_{1/2}$ mSUGRA model
parameter plane for $A_0=0$, $\tan\beta =30$ and $\mu >0$. The red shaded regions are excluded either because the $\ttau_1$ becomes the LSP (in violation 
of bounds from searches for stable charged relics from the Big Bang) or due to
lack of REWSB. The yellow shaded region is excluded by LEP2 searches for 
chargino pair production, and the region below the yellow contour is 
excluded by LEP2 searches for a SM-like Higgs boson, assuming that 
this bound applies to $h$. 
We evaluate the neutralino relic density using the DarkSUSY
program\cite{darksusy} interfaced to Isasugra.
Most of the remaining parameter space is excluded by the 
WMAP bound on $\Omega_{\tz_1}h^2$. The exceptions occur in the blue 
shaded region, where $0.095<\Omega_{\tz_1}h^2<0.129$, and where neutralinos 
can make up {\it all} the CDM of the universe, or in the green shaded region, 
where $\Omega_{\tz_1}h^2<0.095$, and additional CDM candidate(s) 
must exist to saturate the WMAP value of $\Omega_{CDM}h^2$.
The thin region at low $m_0$ along the boundary of the excluded region
is the stau co-annihilation corridor. The region at low $m_0$ and 
low $m_{1/2}$ is the bulk annihilation region, and generally gives rise 
to a Higgs boson mass in conflict with bounds from LEP2 searches.
There is a remaining green region at low $m_{1/2}$ at the edge of the 
LEP2 excluded region where neutralinos can efficiently annihilate through
the light Higgs pole: $2m_{\tz_1}\simeq m_h$. Not shown is the 
HB/FP region at very large $m_0$, since we want to focus 
on the stau co-annihilation region in this section.
\FIGURE{\epsfig{file=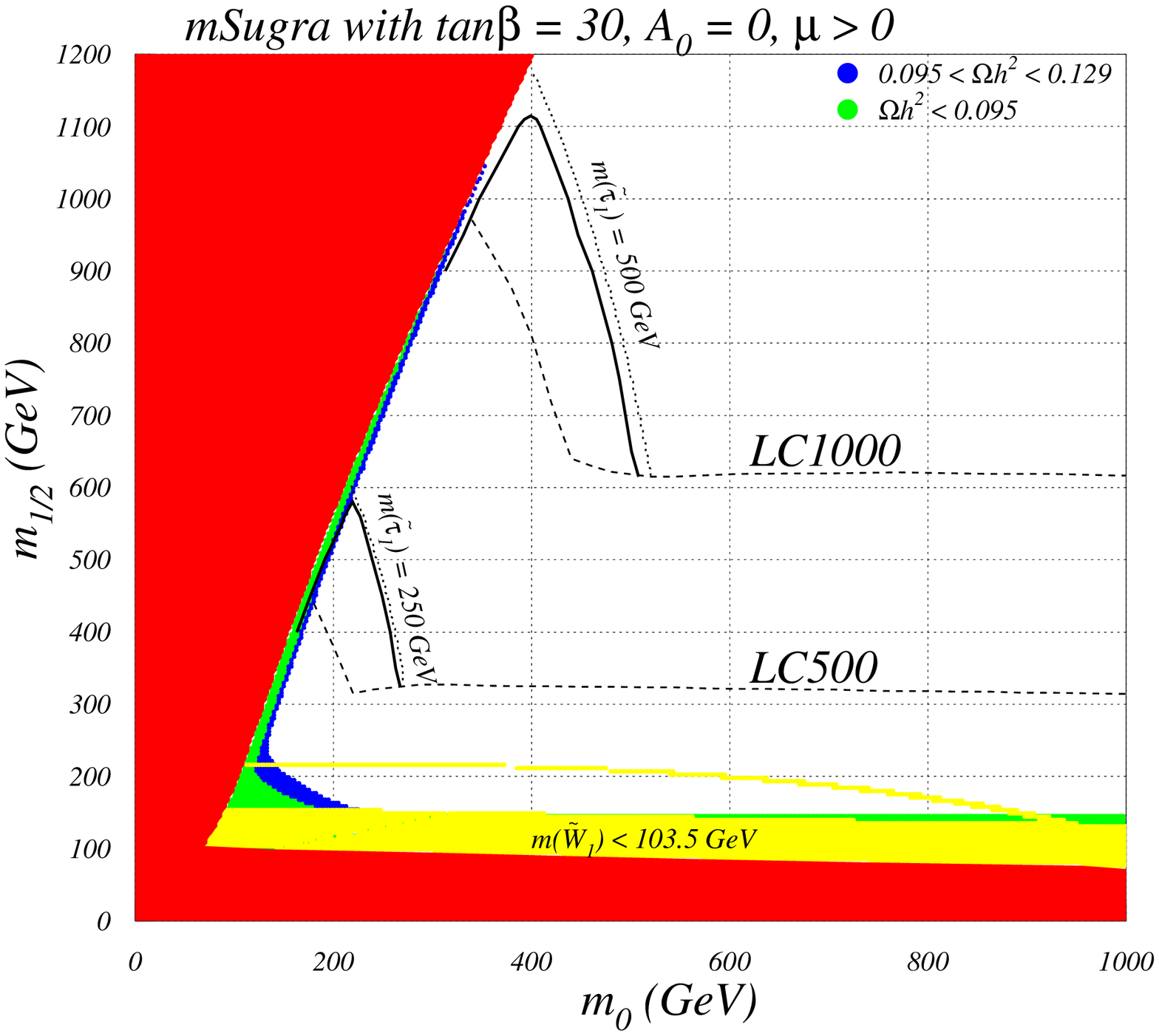,width=14cm} 
\vspace*{-0.8cm}
\caption{Reach of a linear collider for supersymmetry in the mSUGRA
model for $\sqrt{s}=500$ and $1000$~GeV in the stau co-annihilation region,
for $\tan\beta =30$, $A_0=0$ and $\mu >0$. 
The reach via selectron, smuon and chargino pair 
searches is denoted by the dashed contours, while the additional reach
due to stau pair production is denoted by the solid contour. The dotted contour
denotes the kinematic limit for stau pair production.
The red region is 
theoretically excluded, while the yellow region 
is excluded by LEP2 measurements. Below the yellow contour, $m_h \leq
114.4$~GeV. The blue region is within the WMAP $\Omega_{\tz_1}h^2$
$2\sigma$ limit, while the green region has $\Omega_{\tz_1}h^2$ below
the WMAP $2\sigma$ limit.
}
\label{fig:30p}
}

The dashed contours in Fig. \ref{fig:30p} denote the reach of a 
$\sqrt{s}=500$ GeV or 1000 GeV $e^+e^-$ LC for SUSY as calculated in
Ref. \cite{bbkt}. 
For the $\sqrt{s}=1000$ GeV runs, we use $\Upsilon =0.29$ and 
$\sigma_z=0.11$ mm.
The horizontal portion of the contour denotes the 
upper limit of parameter space which is explorable via 
chargino pair searches, while the  rising dashed contour at 
lower $m_0$ values gives the reach in mSUGRA
due to selectron and smuon searches. The dotted contours denote the 
kinematic limit for stau pair production at a 500 or 1000 GeV LC.
The solid contour marks the boundary
of the added region where there is a $5\sigma$ signal
for di-tau jet events, assuming
100 fb$^{-1}$ of integrated luminosity. From these contours, we see
that a linear $e^+e^-$ collider can see most of the additional parameter
space which is accessible to stau pair searches. However, when 
$m_{\ttau_1}\simeq m_{\tz_1}$ along the border of parameter space, the tiny
mass gap $m_{\ttau_1}-m_{\tz_1}$ yields very low visible energy, and the
search contours turn over. 
Thus, in the $\tan\beta =30$ case shown, a direct search for stau pair
production allows only a portion of the WMAP allowed region to be covered.
We note here that we have also investigated stau pair signals
consisting of an isolated lepton plus a single tau jet. In this case,
no additional reach was obtained, and in some cases the reach is even
diminished,
due to the large backgrounds (mainly from $WW$ events) 
in the $\ell+\tau$-jet channel.
We also note here that we have investigated the lower $\tan\beta =10$ value.
In this case, $m_{\ttau_1}\simeq m_{\te_R}$, and hardly any reach is gained
by looking for SUSY in the tau pair channel, as opposed to the 
dielectron or dimuon channels.

In Fig. \ref{fig:45m}, we show the same $m_0\ vs.\ m_{1/2}$ plane, except 
we now take $\tan\beta =45$ and $\mu <0$. The shaded regions and contours
denote the same constraints as in Fig. \ref{fig:30p}. The most noteworthy 
point is that the WMAP allowed region has greatly expanded compared
to the $\tan\beta =30$,  $\mu >0$ case shown in Fig. \ref{fig:30p}. This is 
because the $A$ annihilation funnel has moved into the parameter space shown,
and overlaps with the stau co-annihilation region. We can also see that in 
this case, the additional reach gained by the ditau-jet signal is
considerable, and covers a rather large swath of the WMAP allowed region, for
both a 500 GeV LC as well as for a 1000 GeV LC.
\FIGURE{\epsfig{file=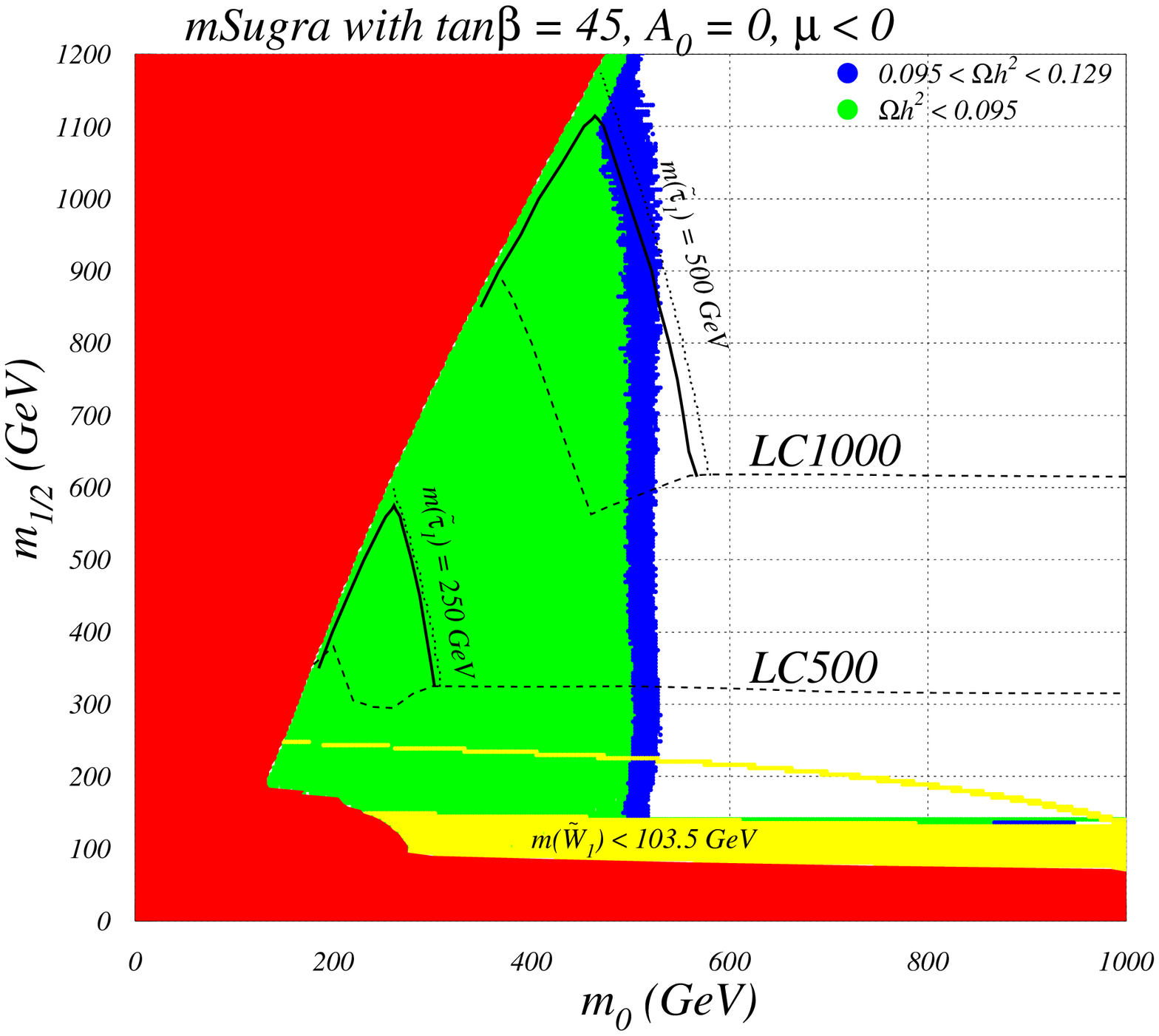,width=14cm} 
\vspace*{-0.8cm}
\caption{Reach of a linear collider for supersymmetry in the mSUGRA
model for $\sqrt{s}=500$ and $1000$~GeV in the stau co-annihilation region,
for $\tan\beta =45$, $A_0=0$ and $\mu <0$. 
The contours and shadings are as in Fig. \ref{fig:30p}.
}
\label{fig:45m}
}

In Fig. \ref{fig:52p}, we again show the mSUGRA model $m_0\ vs.\ m_{1/2}$ 
plane, but now for $\tan\beta =52$ and $\mu >0$. In this case, the 
excluded region where $\ttau_1$ is the LSP has greatly expanded, owing to
the large $\tau$ Yukawa coupling and large left-right mixing, both of which 
act to reduce the $\ttau_1$ mass relative to $m_{\te_R}$ and $m_{\tmu_R}$.
In this plot, the $A$ annihilation funnel is not apparent 
(it would be located in the low $m_0$ excluded region) , 
but its effect is felt over much of the parameter space by adding to
(via off-shell $A$ and $H$ annihilations) the stau annihilation and bulk 
annihilation cross sections, and thus enlarging these regions.
We can see that in this case, the additional reach gained by the 
stau-pair search allows a $\sqrt{s}=500$ GeV $e^+e^-$ collider to
fully access the bulk annihilation region at low $m_0$ and low $m_{1/2}$, 
which would otherwise not be accessible to searches for first 
and second generation sleptons. In addition, some portion of the stau 
co-annihilation region can also be explored by the di-tau signal, although
the region with nearly degenerate $\ttau_1$ and $\tz_1$ remains 
inaccessible due to the tiny mass gap, and low visible energy release
from tau slepton decays. Likewise, the $\sqrt{s}=1000$ GeV LC can
explore all this and more, including a substantial chunk of the 
stau co-annihilation corridor, provided the mass gap between 
$\ttau_1$ and $\tz_1$ is large enough to yield a sufficient rate 
for observable signals after cuts. 
\FIGURE{\epsfig{file=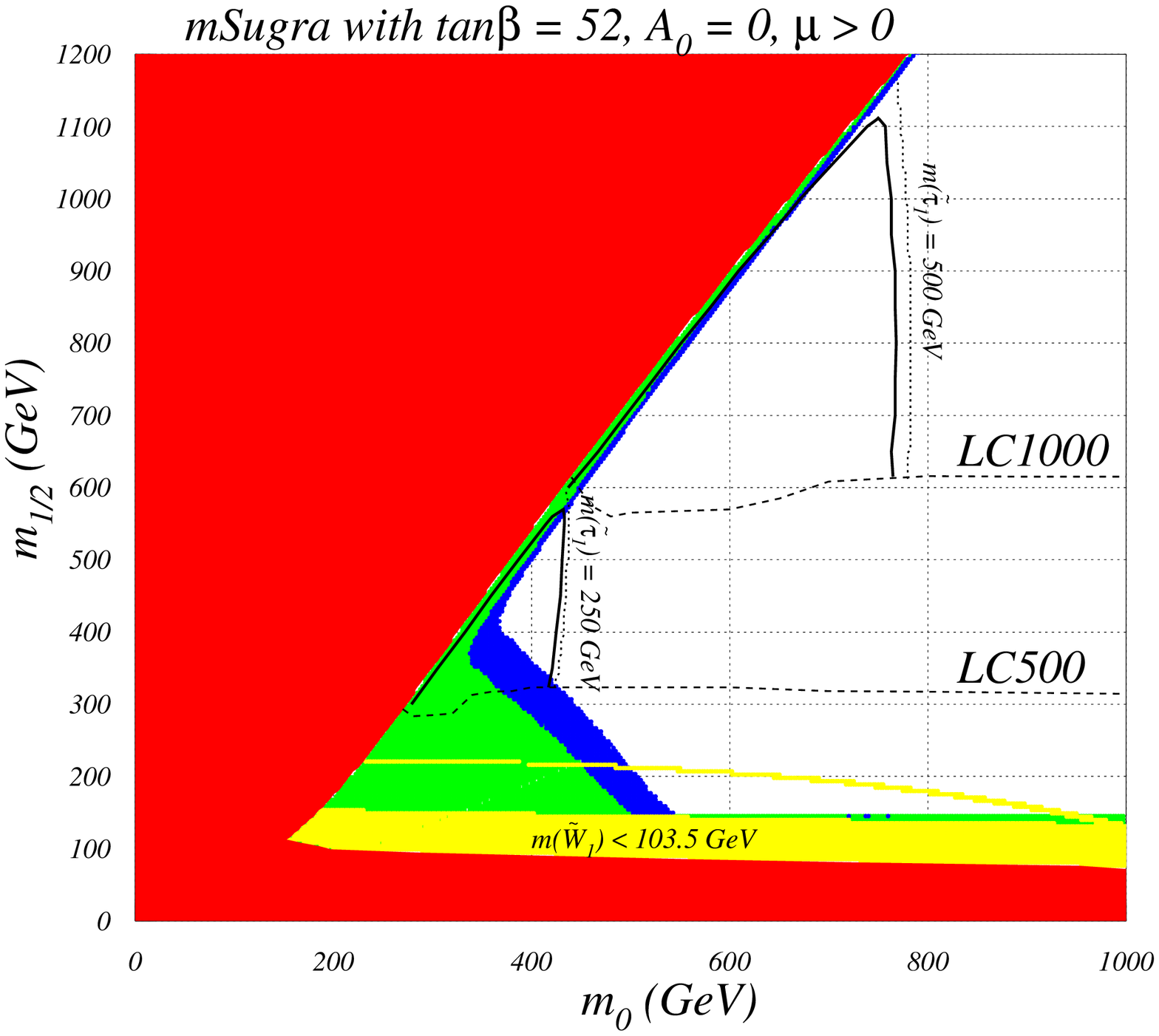,width=14cm} 
\vspace*{-0.8cm}
\caption{Reach of a linear collider for supersymmetry in the mSUGRA
model for $\sqrt{s}=500$ and $1000$~GeV in the stau co-annihilation region,
for $\tan\beta =52$, $A_0=0$ and $\mu >0$. 
The contours and shadings are as in Fig. \ref{fig:30p}.
}
\label{fig:52p}
}

We note in passing that in the stau coannihilation region the reach of
the LHC exceeds that of even the 1~TeV LC \cite{bbkt}.

\section{Reach in the HB/FP region for $m_t=180$ GeV}
\label{sec:hb_fp}

In Ref. \cite{bbkt}, chargino pair production was examined against 
SM backgrounds using a set of standard cuts first suggested in 
Ref.~\cite{jlc1}.
Briefly, the standard chargino pair cuts are as follows.
Following Refs. \cite{jlc1} and \cite{bmt}, 
it was required
to have one isolated lepton plus two jets with, 
{\it i}) 20~GeV $<E_{vis}<\sqrt{s}-100$~GeV, {\it ii}) if $E_{jj}>200$~GeV, 
then $m(jj)<68$~GeV, {\it iii}) $E_T^{mis}>25$~GeV, {\it iv}) 
$|m(\ell \nu )-M_W|>10$~GeV for a $W$ pair hypothesis, 
{\it v}) $|\cos\theta(j)|<0.9$, $|\cos\theta (\ell )|<0.9$,
$-Q_\ell\cos\theta_\ell <0.75$ and $Q_\ell\cos\theta (jj)<0.75$, 
{\it vi}) $\theta_{acop}(WW)>30^\circ$, for a $W$ pair hypothesis.
The reach for $1\ell +2j +\eslt $  events from chargino pair
production was evaluated using a left polarized beam with 
$P_L=+0.9$. These cuts worked well for almost all of parameter
space, save for the large $m_{1/2}$ portion of the HB/FP region.
In that region, the $\tw_1 -\tz_1$ mass gap becomes so small that there
is very little energy release in chargino pair production events, and
the signal would usually fail the standard cuts. A specialized set of cuts
was proposed in Ref. \cite{bbkt} to access the signal
in the large $m_{1/2}$ portion of the HB/FP
region. These  require:
\begin{itemize}
\item 1 isolated lepton plus two jets,
\item 20 GeV$<E_{visible}<100$ GeV,
\item $\cos\phi (jj)>-0.6$,
\item $m(\ell,j_{near})>5$ GeV.
\end{itemize}
The cut on $\cos\phi (jj)$ is on the dijet opening angle in the 
transverse plane. Backgrounds from $2\to 2$ processes were evaluated with 
Isajet, while $\gamma\gamma \to f\bar{f}$ background was evaluated using 
Pythia, which included only the bremsstrahlung portion of the photon
PDF. The total SM background from $2\to 2$ 
processes plus $\gamma\gamma \to f\bar{f}$ was found to be 0.97 fb for a 
$\sqrt{s}=500$ GeV LC. Using these  cuts, it was 
found that the chargino pair signal could be seen essentially up to 
the kinematic limit even in the high $m_{1/2}$ portion of the HB/FP
region where there is no observable signal at the LHC.

In this section, we re-examine the chargino pair signal, but this
time including as well the $\gamma\gamma\to c\bar{c}$ and $b\bar{b}$
backgrounds from both bremsstrahlung and beamstrahlung, as incorporated
into Isajet 7.70. The background estimate from $2\to 2$ processes plus
$\gamma\gamma\to f\bar{f}$ obtained for a $\sqrt{s}=500$
GeV LC with $\Upsilon= 0.1072$ and $\sigma_z =0.12$ mm is now 44.1 fb, after
the above cuts are applied - an increase by a factor of $\sim 40$! (The
background level using standard cuts hardly changes after including
beamstrahlung photons, since that set of cuts is also adept at removing
the $\gamma\gamma$ background.)

The situation is illustrated in Fig. \ref{fig:distctj}, where we show
the distribution in $\cos\theta (j)$ for the most energetic jet in
signal and background events which pass the specialized HB/FP cuts.
The mSUGRA model signal point is taken to be $m_0=8850$ GeV, 
$m_{1/2}=400$ GeV, $A_0=0$, $\tan\beta =30$ and $\mu >0$, where we
have adopted the value $m_t=180$ GeV for the top  quark mass. It is seen
from Fig. \ref{fig:distctj} that the signal events have a broad 
distribution in $\cos\theta (j)$, while the background distribution
is sharply peaked at $|\cos\theta (j)|\sim 1$. 
Considerable suppression of the $\gamma\gamma \to c\bar{c},\ b\bar{b}$ 
background at little cost to signal can thus be gained by requiring in 
addition
\begin{itemize}
\item $|\cos\theta (j)| <0.8$ ,
\end{itemize}
which we apply to both jets in the $\ell +2-jet$ events.
After applying this cut, the $2\to 2$ plus $\gamma\gamma$ background level 
drops from 44.1 fb to 0.39 fb. The signal cross section for the parameter space
point shown drops from 39.5 fb to 28.8 fb.
We have re-evaluated the reach projections in the $m_0\ vs.\ m_{1/2}$
plane of Ref. \cite{bbkt} after including the updated background including 
beamstrahlung and the additional $|\cos\theta (j)|$ cut, and find 
that the reach projections suffer no visible change.
\FIGURE{\epsfig{file=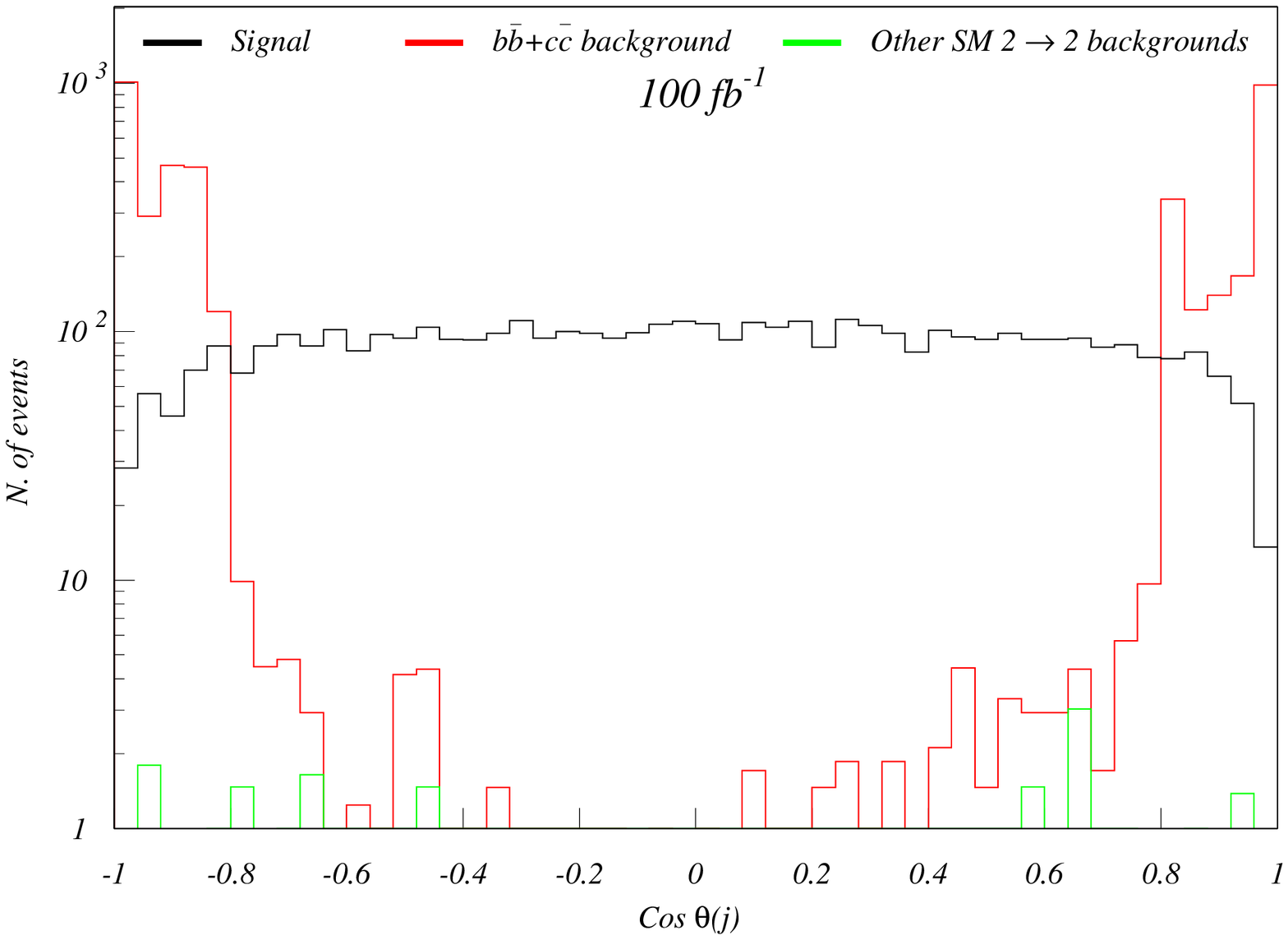,width=14cm} 
\vspace*{-0.8cm}
\caption{The angular distribution of the hardest jet 
in 
$\ell +jj+\emiss$ events from sparticle pair production
at mSUGRA point $m_0,\ m_{1/2},\ A_0,\ \tan\beta,\ sign(\mu )=
8850,\ 400,\ 0,\ 30,\ +1$, with $m_t=180$ GeV for a $\sqrt{s}=500$ GeV LC.
The relic density for this point is $\Omega_{\tz_1}h^2=0.09$.
We show background from $b\bar{b}$ and $c\bar{c}$ production via both
$e^+e^-$ and $\gamma\gamma$ annihilation, along with
background from all other SM processes. 
}
\label{fig:distctj}
}

After the publication of Ref. \cite{bbkt}, the D\O \ Collaboration
has announced a re-analysis of old data on top quark pair production, 
using more sophisticated techniques that retain more information about 
the event shapes for extracting the top quark mass
in the $\ell+jets$ channel\cite{d0mt}. They now find a value 
$m_t=180.1\pm 5.3$ GeV, which has smaller error bars and a 
significantly larger value of $m_t$ than previous measurements.
Combined with CDF measurements, the two experiments find a 
world average $m_t=178.0\pm 4.3$ GeV\cite{cdfd0}.
The value of $m_t$ is an important parameter in determining the 
location of the  HB/FP region in the mSUGRA model\cite{bkt}.
We take this opportunity to update our mSUGRA reach projections
assuming a value of $m_t=180$ GeV, whereas previous results used
$m_t=175$ GeV\cite{bbkt}.

To illustrate the effect of the higher top quark mass, we show
in Fig. \ref{fig:30p_mt180} the $m_0\ vs.\ m_{1/2}$ plane for
$A_0=0$, $\tan\beta =30$ and $\mu >0$. The main effect of the increased value
of $m_t$ is to push the HB/FP region from the vicinity of 
$m_0\sim 3-8$ TeV (for $m_t=175$ GeV) to $m_0\sim 8-14$ TeV
(for $m_t=180$ GeV). Another less noticable effect is that the larger 
$m_t$ value increases the relative value of $m_h$ via radiative corrections, 
so that the region of good relic density from neutralino annihilation
via $s$-channel $h$ exchange has been pulled to larger $m_{1/2}$ values.
This light higgs annihilation corridor is shown as a discontinuous 
narrow band
around $m_{1/2}\sim 150$ GeV in the figure.

We also show the reach of the Tevatron
via the isolated trilepton channel (following procedures listed in
Ref. \cite{bkt}), and the reach of the CERN LHC (following procedures
listed in Ref. \cite{lhc}). The LC reach is plotted as in Ref. \cite{bbkt}.
The Tevatron reach contour here is plotted assuming a $5\sigma$ signal with
10 fb$^{-1}$ of integrated luminosity. It remains qualitatively similar to the
$\tan\beta =30$, $m_t=175$ GeV case, except that it is effectively stretched
out in $m_0$. The light Higgs annihilation corridor, which was not apparent
for $\tan\beta =30$, $m_t=175$ GeV, is now visible, and within the reach of
the Fermilab Tevatron.
The LHC reach contour, plotted for a $5\sigma$ signal with 100 fb$^{-1}$
of integrated luminosity, extends to $m_{1/2}\sim 1300$ GeV for low
$m_0$ (corresponding to a value of $m_{\tg}\sim 3$ TeV). As $m_0$ increases,
the contour drops until $m_0\sim 3$ TeV is reached, whereupon it levels out and stays roughly constant in $m_{1/2}\sim 700$ GeV (corresponding to a
value $m_{\tg}\sim 1.8$ TeV). What is happening is that as $m_0$ increases,
squarks become increasingly heavy, while $m_{\tg}$ remains roughly
constant for a given value of $m_{1/2}$. Thus, at $m_0\sim 3$ TeV, 
the squark contribution to LHC signals has essentially decoupled, and 
almost all the SUSY signal originates from gluino pair production, followed
by gluino cascade decays.
The LC reach plots are qualitatively similar to those presented in 
Ref. \cite{bbkt}, except that they are stretched out in $m_0$ until the
HB/FP region is reached. As expected, this reach is mainly governed by
the chargino mass and the mass gap between the chargino and the LSP.
We note here that the shift in the HB/FP region to very large values
of $m_0$ for $m_t=180$ GeV can be viewed favorably for SUSY theories, 
in that the large scalar masses will give further suppression to 
possible flavor changing or CP violating processes\cite{dine,fmm}.

%
\FIGURE{\epsfig{file=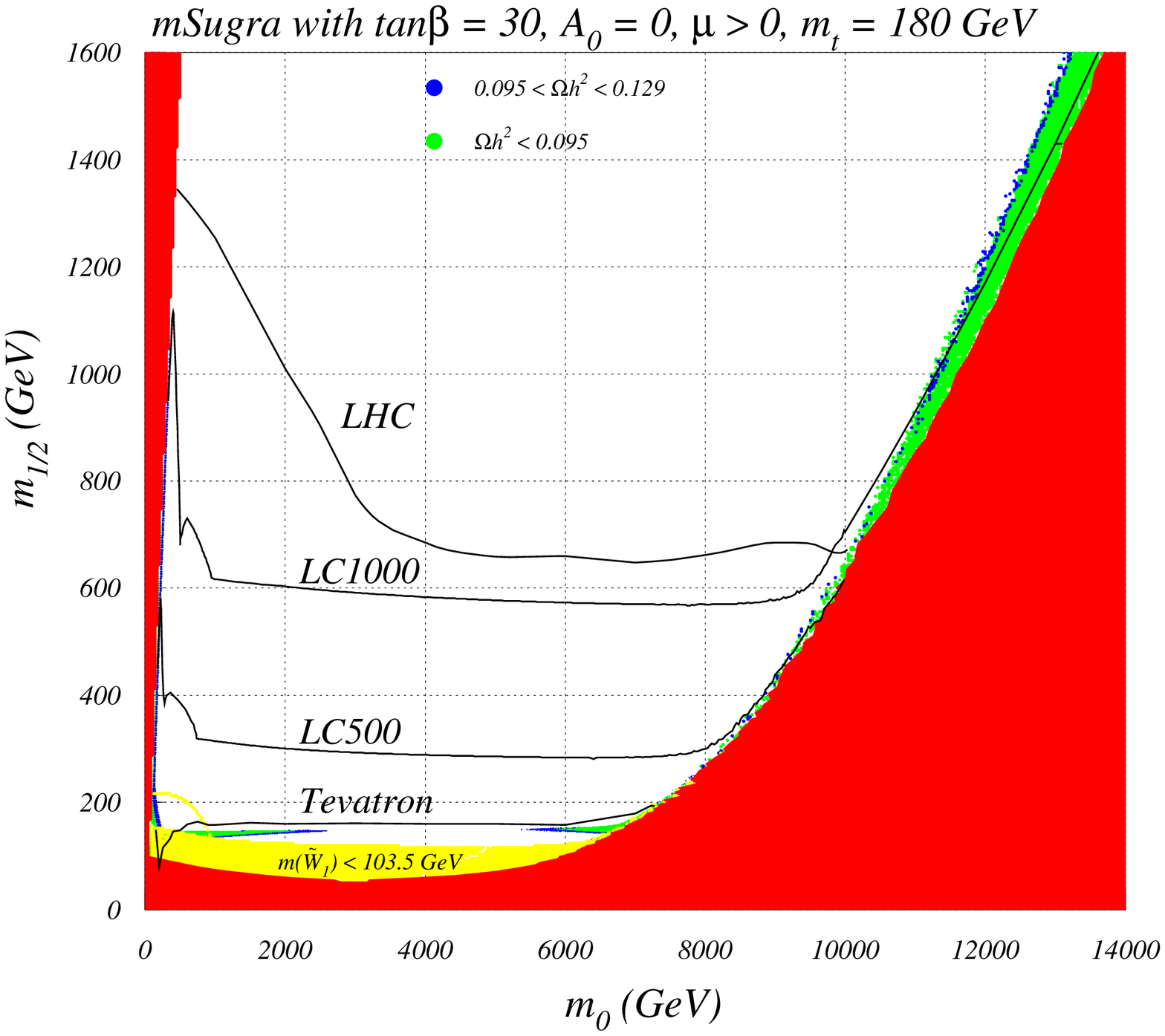,width=14cm} 
\vspace*{-0.8cm}
\caption{Reach of the Fermilab Tevatron, CERN LHC and 
a linear collider for supersymmetry in the mSUGRA
model for $\sqrt{s}=500$ and $1000$~GeV
for $\tan\beta =30$, $A_0=0$, $\mu >0$ and $m_t=180$ GeV. 
The red region is 
theoretically excluded, while the yellow region 
is excluded by LEP2 measurements. Below the yellow contour, $m_h \leq
114.4$~GeV. The blue region is within the WMAP $\Omega_{\tz_1}h^2$
$2\sigma$ limit, while the green region has $\Omega_{\tz_1}h^2$ below
the WMAP $2\sigma$ limit.
}
\label{fig:30p_mt180}
}

\section{Conclusions}
\label{sec:conclude}
%

Significant portions of the mSUGRA parameter space where the neutralino relic 
density is in accord with WMAP analyses occur around the boundaries of
parameter space in the mSUGRA region. These boundaries occur where
$m_{\ttau_1}-m_{\tz_1}\to 0$ in the stau co-annihilation region
or where $m_{\tw_1}-m_{\tz_1}\to 0$ in the HB/FP region. In these
regions, pair production of NLSPs will result in collider
events with very low energy release, and low invariant mass.
The process $\gamma\gamma\to f\bar{f}$ where the $\gamma$s come from
initial state bremsstrahlung or beamstrahlung is also highly peaked at low
$f\bar{f}$ invariant mass, and may thus constitute an important 
background process in these DM allowed regions of parameter space.

We have incorporated $\gamma\gamma\to f\bar{f}$ into the Isajet event generator
for $e^+e^-$ collisions. This enables us to evaluate signal and background 
rates for $e^+e^-\to\ttau_1^+\ttau_1^-\to \tau\bar{\tau} +\eslt$
production in the stau co-annihilation region. Using suitable cuts to
tame the $\gamma\gamma\to\tau\bar{\tau}$ background, we find that significant
regions of additional reach are obtained, especially if $\tan\beta $ is
large.

We have also investigated the $\gamma\gamma \to c\bar{c},\ b\bar{b}$ 
background to $e^+e^-\to\tw_1^+\tw_1^-$$\to \ell+$$jj +$$\eslt$ 
production in the HB/FP region of the mSUGRA model. Inclusion of the 
beamstrahlung backgound significantly increases the SM background 
for this signal. However, the additional background can be effectively 
removed by requiring the jets to be away from the beam directions.

We update our mSUGRA reach plots for $\tan\beta =30$ by increasing the value
of $m_t$ from 175 GeV to 180 GeV, in accord with recent $D\O$ \ measurements.
This has the effect of pushing the HB/FP region out to much large
values of $m_0$, and effectively stretching out the reach contours for
the Fermilab Tevatron, the CERN LHC and an $e^+e^-$ LC. Expressed in
terms of the chargino mass, the reach in the HB/FP region is
qualititatively unaltered.

\acknowledgments

We thank M. Drees for providing us with his subroutines which calculate
beamstrahlung distributions. 
This research was supported in part by the U.S. Department of Energy
under contract numbers DE-FG02-97ER41022 and DE-FG03-94ER40833.
	
%

\end{document}